\documentclass[sigconf, nonacm]{acmart}

\newcommand\vldbyear{2026}
\newcommand\vldbworkshop{Fourth International Workshop on Composable Data Management Systems}
\newcommand\vldbauthors{\authors}
\newcommand\vldbtitle{\shorttitle} 
\newcommand\vldbavailabilityurl{https://github.com/BauplanLabs/long_abstracts_and_other_stories}
\newcommand\vldbpagestyle{empty} 

\usepackage{array}
\usepackage{listings}
\floatstyle{ruled}
\newfloat{listing}{tb}{lst}{}
\floatname{listing}{Listing}

\begin{document}
\title{Not Your Usual Type(s)}
\subtitle{Data contracts as types across languages and engines}
\author{Aldrin Montana}
\authornote{All authors contributed equally. JT is the corresponding author and PI on the project:  \url{mailto:jacopo.tagliabue@bauplanlabs.com}.}
\affiliation{%
 \institution{Bauplan Labs}
 \country{USA}
}

\author{Colin Marc}
\authornotemark[1]
\affiliation{%
 \institution{Bauplan Labs}
 \country{Germany}
}

\author{Luca Bigon}
\authornotemark[1]
\affiliation{%
 \institution{Bauplan Labs}
 \country{Italy}
}

\author{Jacopo Tagliabue}
\authornotemark[1]
\affiliation{%
 \institution{Bauplan Labs}
 \country{USA}
}

\begin{abstract}
Composable data systems promise to let developers combine languages, engines, and catalogs without sacrificing a coherent user experience. In practice, however, pipeline-node boundaries remain weakly specified: transformations exchange tables through schemas that are often checked late, enforced unevenly across languages, and disconnected from the semantics business users care about. Based on over a year of operating millions of jobs in Bauplan, we share the design principles behind our new SDK, which treats data contracts as types for a composable, multi-language lakehouse. Users, whether humans or agents, annotate input and output tables with schema objects that encode column types, constraints, documentation, and lineage; Bauplan then interprets these annotations at different points in the execution lifecycle. We show how this design addresses common production failures, and how an ``everything-as-code'' philosophy enables both deterministic and non-deterministic reasoning over data flows across languages and engines.
\end{abstract}

\maketitle

\pagestyle{\vldbpagestyle}
\begingroup\small\noindent\raggedright\textbf{VLDB Workshop Reference Format:}\\
\vldbauthors. \vldbtitle. VLDB \vldbyear\ Workshop: \vldbworkshop.\\ 
\endgroup
\begingroup
\renewcommand\thefootnote{}\footnote{\noindent
This work is licensed under the Creative Commons BY-NC-ND 4.0 International License. Visit \url{https://creativecommons.org/licenses/by-nc-nd/4.0/} to view a copy of this license. For any use beyond those covered by this license, obtain permission by emailing \href{mailto:info@vldb.org}{info@vldb.org}. Copyright is held by the owner/author(s). Publication rights licensed to the VLDB Endowment. \\
\raggedright Proceedings of the VLDB Endowment. 
ISSN 2150-8097. \\
}\addtocounter{footnote}{-1}\endgroup

\ifdefempty{\vldbavailabilityurl}{}{
\vspace{.3cm}
\begingroup\small\noindent\raggedright\textbf{VLDB Workshop Artifact Availability:}\\
The source code, data, and/or other artifacts have been made available at \url{https://github.com/BauplanLabs/long_abstracts_and_other_stories}.
\endgroup
}

\section{Introduction} \label{sec:intro}

The data lakehouse is today the standard architecture for analytics and AI workloads, combining object storage with open table formats and decoupled, multi-language compute \cite{Zaharia2021LakehouseAN,mazumdar2023datalakehousedatawarehousing}. In many lakehouses, data pipelines -- DAGs of transformations from raw to refined data assets -- are the most common OLAP use case, measured by both usage and total cost \cite{cdms2025eudoxia,Renen2024}. 

Historically, a large fraction of DAG errors are due to schema changes at the intersection of two nodes \cite{FOIDL2024111855}, as columns get dropped or replaced, types change, and semantics shift. The problem is even more acute in composable data systems \cite{10.14778/3603581.3603604}, where multi-language, multi-runtime pipelines are the norm and the node interface may be fully unconstrained or unevenly enforced across languages and runtimes. As the majority of labor shifts from writing to verifying code, thanks to the rise of agents \cite{huang_control_2025}, self-documenting, ``correct-by-design'' data projects promise to bring to data engineering the same productivity boost experienced by software engineers \cite{tagliabue2025safeuntrustedproofcarryingai}.

As most data correctness errors are \textit{prima facie} the same avoidable runtime errors that plague dynamic languages \cite{sun2025coevolutiontypesdependenciesrepositorylevel}, we redesigned our pipeline SDK by borrowing a familiar solution from dynamic languages: type annotations \cite{10.1145/3540250.3549114}. In \textit{this} paper, we present the design of the new \textbf{Bauplan} SDK (``SDK 2.0''), which defines a contract layer for composable lakehouse DAGs. This layer makes schemas, constraints, lineage, and documentation portable across languages, execution engines, and lifecycle stages. Once contracts are explicit and machine-checkable, both humans and agents can safely operate over the same dataflow interface. In particular, we summarize our contributions as follows:

\begin{enumerate}
  \item \textbf{multi-language DAG types}: thanks to our privileged position running Bauplan -- an agentic lakehouse at scale with millions of DAGs in production -- we motivate SDK 2.0 through real-world failures, and outline how the new abstractions address them;
  \item \textbf{three-stage contract enforcement}: we map contract validation to three distinct moments in pipeline execution across a distributed platform: local inference over code files, planning inference over lakehouse state, and runtime inference over data assets produced in a run. As with software engineering, our guiding principle for agentic data engineering is to ``shift left'' the burden of correctness and fail as early as possible with a clear, explicit error message;
  \item \textbf{everything-as-code}: table descriptions, downstream metrics, joinable columns and column metadata all \textit{live next to the code} that generates the table, and persist into the table metadata. Data semantics (i.e. how tables connect to business entities) are essential both for ongoing maintenance and for translation of business questions into analytics code \cite{9140254}: by ``shifting left'' the burden of documentation, we enable a full agentic loop on the code base over a longer time horizon.
\end{enumerate}

Altogether, these capabilities continue the convergence of data and software engineering, which we, among others, pioneered before the explosion of coding agents. While \textit{Bauplan} is the natural application ground for these ideas, our design sits more generally at the intersection of programming languages, data management and distributed systems. As such, the ergonomics introduced here are easily portable to other composable systems and the guarantees are easily generalizable to other runtimes. All in all, we believe our lessons from the trenches to be valuable to a broad set of data practitioners.

\section{Platform design and failure modes} 
\label{sec:platform}

\subsection{Bauplan overview} 
\label{sec:bauplan}

\texttt{Bauplan} is a composable data system built for AI agents as first-class citizens \cite{pre:sheng-correctdesign,tagliabue2025safeuntrustedproofcarryingai,tagliabue2026queryingoncesupervaluationismagentic}. It is \textit{composable} as it re-uses a mix of proprietary and open source modules \cite{bigon2025dag,curtin2025deconstructed} assembled with clear interfaces, while offering users a seamless, cohesive experience. In particular, Bauplan is a lakehouse built on the separation of storage (Apache Iceberg \cite{iceberg} on object storage) and compute (SQL and Python sandboxes). While we refer the reader to \cite{10.1145/3702634.3702955,Tagliabue2023BuildingAS,10825377} for a full system overview, we survey here the distinctive features that are relevant for our SDK re-design.

We introduce a running example both to illustrate Bauplan's original SDK and to introduce the platform abstractions. With ``Titanic'' as the source
dataset\footnote{\url{https://www.kaggle.com/datasets/yasserh/titanic-dataset}}, we consider a three-node DAG. Node \texttt{f} is a SQL transformation that produces table ``f'' (through a naming convention), Node \texttt{g} is a Python transformation that produces table ``g'', and Node \texttt{h} is a Python transformation that produces table ``h''. Listings \ref{lst:f_untyped}, \ref{lst:g_untyped}, \ref{lst:h_untyped} show each transformation expressed with the original SDK.

\begin{lstlisting}[
  language=SQL,
  showstringspaces=false,
  columns=fullflexible,
  caption={Node \#1 (f) as a SQL transformation},
  basicstyle=\ttfamily\scriptsize,
  keepspaces=true,
  label=lst:f_untyped
]
-- bauplan: name=f
-- bauplan: materialization_strategy=REPLACE
  SELECT PassengerId AS pid, Pclass AS tclass, Fare
  FROM titanic WHERE Pclass IN (1, 2) AND Sex = 'female'
\end{lstlisting}

\begin{lstlisting}[
  language=Python,
  showstringspaces=false,
  columns=fullflexible,
  caption={Node \#2 (g) as a Python transformation},
  basicstyle=\ttfamily\scriptsize,
  keepspaces=true,
  label=lst:g_untyped
]
@bauplan.python('3.13', pip={'polars': '1.37'})
@bauplan.model(materialization_strategy='REPLACE')
def g(
  adult_lodgings=bauplan.Model(
    'titanic',
    columns=['PassengerId', 'Pclass', 'Fare', 'Cabin'],
    filter='Age >= 18 AND Age <= 65',
  ),
):
  return (
    pl.DataFrame(adult_lodgings)
      .with_columns(cabin_level=(pl.col('Cabin').str.slice(0, 1)))
      .with_columns(mean_fare=(pl.col('Fare').mean().over('cabin_level')))
      .select(pl.col('PassengerId').alias('pid'),
              pl.col('Pclass').alias('tclass'),
              'Cabin', 'cabin_level', 'Fare', 'mean_fare')
      .to_arrow()
  )
\end{lstlisting}

\begin{lstlisting}[
  language=Python,
  showstringspaces=false,
  columns=fullflexible,
  caption={Node \#3 (h) as a Python transformation},
  basicstyle=\ttfamily\scriptsize,
  keepspaces=true,
  label=lst:h_untyped
]
@bauplan.python('3.13', pip={'polars': '1.37'})
@bauplan.model(materialization_strategy='REPLACE')
def h(f_data=bauplan.Model('f', columns=['pid', 'tclass']),
      g_data=bauplan.Model('g')):
    return (pl.DataFrame(f_data)
              .join(pl.DataFrame(g_data), on=('pid', 'tclass'))
              .filter((pl.col('Fare') > pl.col('mean_fare')))
              .select('pid', 'tclass', 'Fare', 'Cabin')
              .to_arrow())
\end{lstlisting}

\begin{figure}
    \centering
    \includegraphics[width=\columnwidth]{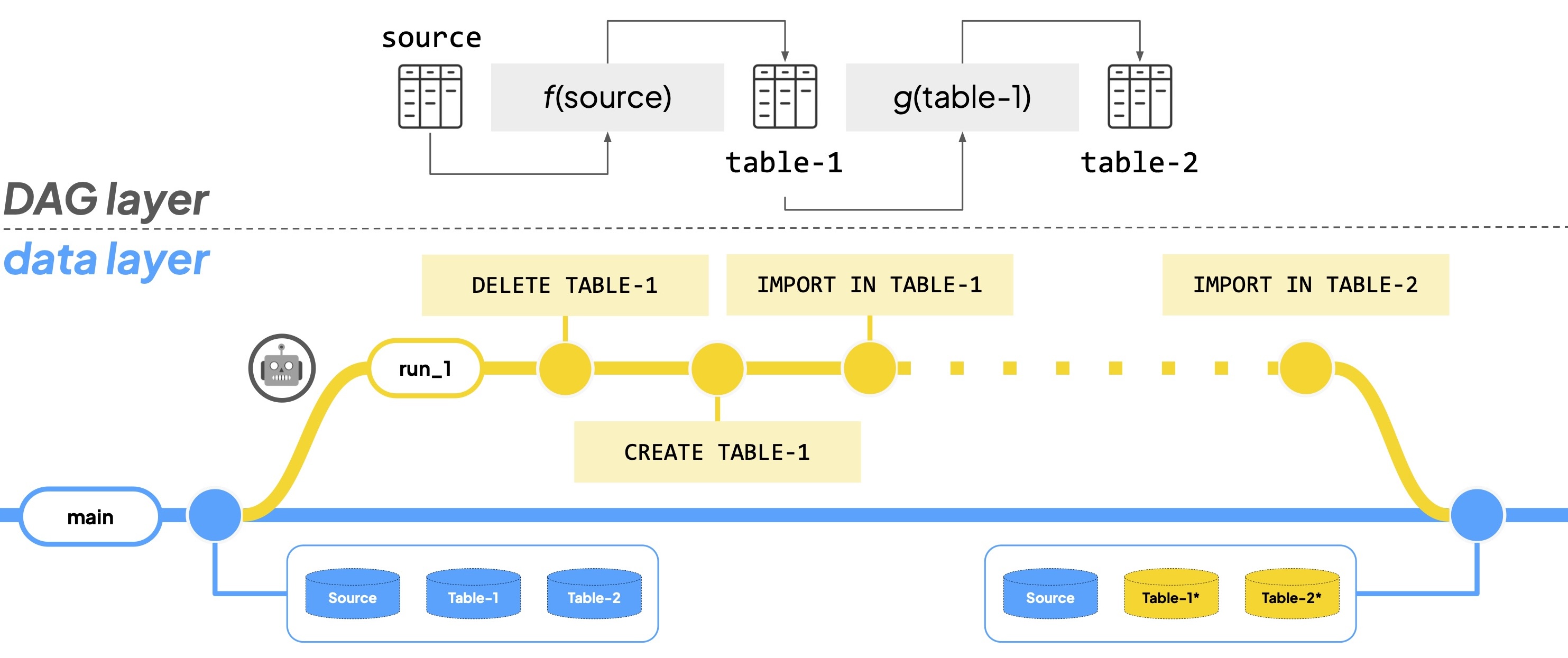}
    \caption{Pipeline declarative code and lakehouse changes. We show a simplified two-node DAG (\textit{top}) transforming data from a source table into \textit{table-1} and \textit{table-2}: from the data asset perspective, the language and logic of the transformations are implementation details for the lakehouse (\textit{bottom}), which is recording on a data branch the I/O operations specified in the code.}
    \label{fig:iso}
\end{figure}

The abstractions are largely self-explanatory, so we highlight only their declarative aspects, which play an important role in enabling ``DAG types'':

\begin{description}
    \item[DAG Topology] Python function parameters and SQL query source tables collectively
    define the DAG topology; \texttt{f} $\rightarrow$ \texttt{h} and \texttt{g} $\rightarrow$
    \texttt{h} for the above snippets. When considering a SQL query as a function
    (as \textit{dbt}\footnote{\url{https://github.com/dbt-labs/dbt-core}} does), it is easy
    to realize that transformations \texttt{f} and \texttt{g} share the same language-independent shape:
one or more input tables and a single output table. Restricting the signature does not significantly constrain pipeline topologies and enables language-agnostic enforcement
    of invariants at the boundaries (Section~\ref{sec:types}).
    \item[Infrastructure] Python version and packages are listed in a decorator
    \textit{per function}, requiring no client-side tools (such as Docker) and no installation
    commands; it is the platform's job to make sure that the sandbox has the required dependencies
    when running the function.
    \item[I/O] Physical data operations (e.g., reading source data from S3 in \texttt{f} and
    \texttt{g}, passing their outputs into \texttt{h}, writing transformed data, and applying
    filter and projection pushdown in \texttt{g})
    happen in ``platform space'', so that user space only contains transformation code that is
    executed in secure and sandboxed compute. Fig.~\ref{fig:iso} (\textit{bottom}) showcases the
    ``isomorphism'' \cite{schneider2026skillissuesdatacentricoptimization} between the declarative layer and the corresponding data changes.
\end{description}

We will refer to this example (and variations on it) in the rest of the paper.

\subsubsection{Execution model} 

\begin{figure}
\centerline{\includegraphics[width=0.45\textwidth]{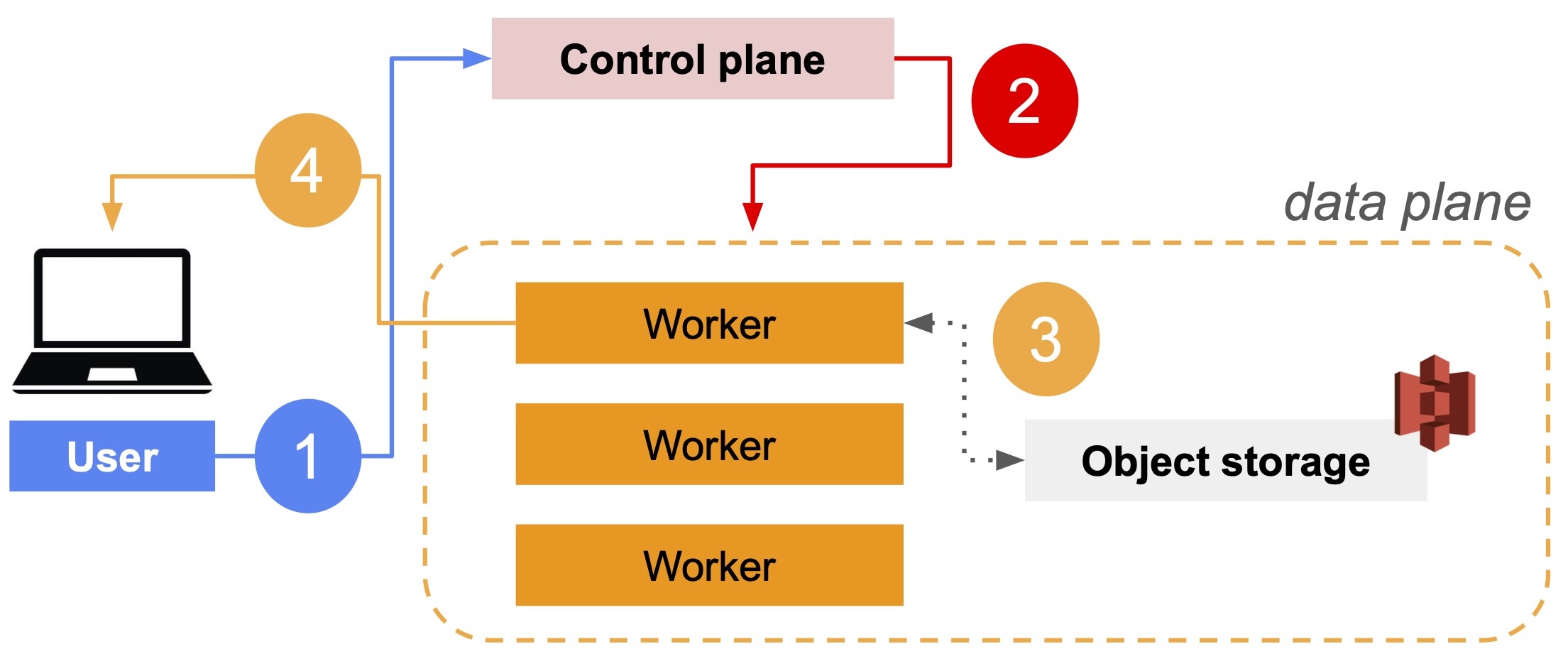}}
\caption{\textbf{A \textit{run} in Bauplan}: 1) a user writes code locally, and triggers the run; 2) the control plane parses the code into a plan and sends it to a worker for execution; 3) the worker reads/writes data from/to S3 and 4) streams logs and results to the user.}
\label{fig:cloud}
\end{figure}

\texttt{Bauplan} has a standard lakehouse architecture
(Figure~\ref{fig:cloud}): a control plane, a FaaS data plane, and a local
client (CLI, SDK) to trigger cloud
operations~\cite{10.1145/3702634.3702955}. Fig.~\ref{fig:cloud} shows the
logical flow of information when a DAG gets executed (a ``run'')---from a user's local environment to object storage, and back. A coding agent writes a Bauplan
project locally and submits the project when triggering the run; the control
plane parses the code into a plan and sends it to a worker for execution;
the worker reads/writes data from/to S3 and streams logs and result
tuples back. The mapping between user code and lakehouse changes is depicted
for a linear DAG in Fig.~\ref{fig:iso}: since tables are managed through
Git-like abstractions \cite{10.1145/3650203.3663335,sheng2026gitlakegitfordataagenticlakehouse}, every function
execution in the DAG corresponds to a \textit{commit}, i.e., an immutable
reference to the state of the lakehouse at that time. For example, since
\texttt{g} is declared with $REPLACE$ semantics, the platform will drop and
re-create the table during a run, then insert rows: what the user writes
as a single function gets (deterministically) ``compiled'' as a series of
physical data operations on the lake.

Importantly, the logical flow of a \textit{run} identifies three key \emph{moments} when executing a DAG: ($1$) the local code environment, before the run is triggered; ($2$) the control plane, when preparing the run (before DAG execution begins); and finally ($3$) the worker process, before and after physical data operations (persisting data is a specific operation). These roughly correspond to code compilation, CI/CD and runtime invariant checks in software development: as a general design principle, we want a system that fails as early as possible and with clear responsibilities at each stage of the process.

\subsection{Failure modes}
\label{sec:failures}

Across millions of production jobs, ``avoidable'' schema mismatches at node boundaries have been a recurring and operationally important class of failure \cite{FOIDL2024111855,bhadauria2026catalogdataerrors}. Note that while these failures are already challenging for workloads run by human experts, their severity is amplified in a world in which untrusted agents operate on production data at scale \cite{tagliabue2025safeuntrustedproofcarryingai}. To illustrate common failure modes, we consider a few scenarios:

\begin{description}
    \item [columns dropped] If $tclass$
    is dropped from \texttt{g} without dropping references to it in
    \texttt{h}, we see an error in \texttt{h} \textit{at runtime}.
    \item [null values present] 
    The $Pclass$ column, and therefore the derived $tclass$ column, contains no $NULL$ values, although both are declared nullable in Iceberg. If \texttt{g} filters out $NULL$ values from its output,
    there is no way to communicate that intent to other DAG nodes:
    constraints cannot be defined, enforced, or even remembered for
    query results.
    \item [type change] If \texttt{g} were to change $tclass$ to have string
    values such as ``First'' for first class instead of the numerical value
    $1$, it would implicitly alter the column type. This would raise an
    error for \texttt{h} when it is used to join with the table ``f''
    \textit{at runtime}.
    \item [shifting semantics] Consider that $tclass$ from
    \texttt{f} is filtered to specific values, whereas $tclass$ from
    \texttt{g} is unfiltered. Downstream analyses can easily look correct
    but actually suffer from errors of omission due to confusing similarly
    named columns in (or after) a join operation. Column lineage can be used
    to disambiguate columns by origin (which DAG node they flow from).
\end{description}

Pushing semantic information to live next to everything else is part of
Bauplan's original vision of ``everything-as-code'', in line with many
modern Pythonic frameworks \cite{Tagliabue2023ReasonableSM}. Additionally, beyond these failures, a second trend has been pushing us to revise SDK 1.0: as downstream consumers of refined assets are themselves agents free to explore the lakehouse, many of the original
semantic layer \cite{rumiantsau2026semanticlayersreliablellmpowered}
capabilities (such as query compilation and lineage) seem ripe for
disintermediation by on-the-fly LLM reasoning.

\section{DAG Types} \label{sec:types}

In this section, we discuss our approach to evolving Bauplan in a principled way to address common challenges rooted in schema failures. Our approach follows a typical two-part pattern from dynamic languages such as Python:
($1$) \textit{optional syntax}---annotations of code for both human and
machine verification and ($2$) \textit{enforcement semantics}---how a
verifier interprets the annotations.

\subsection{Annotation Syntax}

We address hidden, unintentional schema failures by defining syntax for explicit, machine-checkable contracts. This allows developers to
declaratively specify expectations between transformations and incrementally add them to their Bauplan projects to achieve schema and column lineage inference, as exemplified by the snippets below\footnote{Please check the accompanying repository for the full-fledged example: redundant details are omitted here for brevity}.

  \begin{lstlisting}[
    language=Python,
    showstringspaces=false,
    columns=fullflexible,
    caption={Type contract syntax},
    basicstyle=\ttfamily\scriptsize,
    keepspaces=true,
    label=lst:schemas
  ]
  class WomenByClass(TableSchema):
    """Schema for women passengers with 1st or 2nd class tickets."""
    pid:    Annotated[Int64, Required, Doc('Unique passenger ID')]
    tclass: Annotated[Int64, Required,
        Doc('Ticket class, filtered to 1st or 2nd class only.')
    ]
    Fare:   Annotated[Float64, Doc('Ticket cost (British pounds)')]

  class CabinCostSource(TableSchema):
    """Schema for cabin costs used for analysis."""
    # Same annotations as WomenByClass['pid', 'tclass']
    PassengerId: Annotated[Int64, ...]
    Pclass:      Annotated[Int64, ...]
    Fare:  Annotated[Float64, Doc('Ticket cost (British pounds)')]
    Cabin: Annotated[String,  Doc('Cabin number (C28 is on deck C)')]

  class CabinCostAnalysis(TableSchema): ... # Omitted for brevity
  
  class ExpensiveWomenCabins(TableSchema):
    """Schema for expensive cabins reserved by working-age women."""
    # Lineage by referencing schema column (`Schema['col']`)
    pid:    Annotated[Int64,   WomenByClass['pid']]
    tclass: Annotated[Int64,   WomenByClass['tclass']]
    Cabin:  Annotated[String,  CabinCostAnalysis['Cabin']]
    Fare:   Annotated[Float64, WomenByClass['Fare'],
        Doc('Cost of tickets that cost more than average (by deck).')
    ]
  \end{lstlisting}
  
Listing~\ref{lst:schemas} illustrates how schema objects are defined
and how they are associated with rich annotations. Then,
Listings~\ref{lst:f_typed}, \ref{lst:g_typed}, and \ref{lst:h_typed}
show how each node in the DAG directly associates its input tables
(from upstream transformations) and its output table with defined
schema objects. In this way, schema objects elegantly relate documentation and constraints to DAG nodes using a lightweight, extensible mechanism.

\textbf{Schema Objects} are defined using a Python-centric design
to minimize the need for custom SQL and because we expect the syntax
to be more ergonomic in Python.\footnote{Note that while the design does not require a specific version of Python, we show syntax best supported in Python $>=3.13$ for convenience} This means that we lean heavily into the use of standard type annotations using $typing.Annotated$ and familiar Pydantic-style annotations \cite{gh:pydantic} (type objects such as $Required$ and classes such as $Doc$ for objects that don't support docstrings).

Bauplan uses the base class $TableSchema$ to identify Bauplan schema objects and uses the special $Annotated$ type to identify column annotations. The first annotation is the column's data type and subsequent annotations may be rich metadata for ``semantic layer'' support (Section~\ref{sec:semanticlayer}) or may be constraints for validation (Section~\ref{sec:quality}). For column data types, we choose to accept generic names for supported types (such as $String$ and $Float64$) and Bauplan defines a direct correspondence to PyArrow types: to maximize compatibility with other platforms (Section~\ref{sec:compo}), we support the subset of Arrow types that intersects with Iceberg. To identify explicit column lineage, a schema column may be referenced directly with the syntax $Schema['column\_name']$. For example, the annotation $WomenByClass['pid']$ on $ExpensiveWomenCabins.pid$ explicitly records lineage to the $pid$ field in the $WomenByClass$ schema, allowing the system and downstream tools to recover the corresponding upstream origin and metadata.

  \begin{lstlisting}[
    language=SQL,
    showstringspaces=false,
    columns=fullflexible,
    caption={Node \#1 (f) with output schema reference},
    basicstyle=\ttfamily\scriptsize,
    label=lst:f_typed
  ]
  -- bauplan: name=f
  -- bauplan: materialization_strategy=REPLACE
  -- bauplan: output_schema=WomenByClass
  SELECT PassengerId AS pid, Pclass AS tclass, Fare
  FROM titanic WHERE Pclass IN (1, 2) AND Sex = 'female'
  \end{lstlisting}

  \begin{lstlisting}[
    language=Python,
    showstringspaces=false,
    columns=fullflexible,
    caption={Node \#2 (g) with typing},
    basicstyle=\ttfamily\scriptsize,
    label=lst:g_typed
  ]
  @bauplan.python('3.13', pip={'polars': '1.37'})
  @bauplan.model(materialization_strategy='REPLACE')
  def g(
      adult_lodgings: Annotated[
          Table[CabinCostSource],
          Filter(Model('titanic'), 'Age >= 18 AND Age <= 65')
      ],
  ) -> Table[CabinCostAnalysis]:
    """Analyze the cabin cost by level for working-age adults."""
    # Function body remains unchanged
    return (pl.DataFrame(adult_lodgings).(...).to_arrow())
  \end{lstlisting}

  \begin{lstlisting}[
    language=Python,
    showstringspaces=false,
    columns=fullflexible,
    caption={Node \#3 (h) with typing},
    basicstyle=\ttfamily\scriptsize,
    keepspaces=true,
    label=lst:h_typed
  ]
  @bauplan.python('3.13', pip={'polars': '1.37'})
  @bauplan.model(materialization_strategy='REPLACE')
  def h(
      f_data: Annotated[Table[WomenByClass],     Model('f')],
      g_data: Annotated[Table[CabinCostAnalysis], Model('g')],
  ) -> Table[ExpensiveWomenCabins]:
    """Expensive cabins reserved by/for working-age women."""
    # Function body remains unchanged
    return (pl.DataFrame(f_data).(...).to_arrow())
  \end{lstlisting}

\textbf{Transformation Annotations} are defined by parameter and return type annotations in Python and by a structured comment in SQL. In Python, a function parameter is annotated using $Annotated$ where the first annotation is a $Table[schema]$ type, meaning a tabular data structure (such as $pyarrow.Table$) whose schema matches the specified schema contract. The second annotation must resolve to a $Model$ reference identifying an input transformation (an Iceberg table or DAG node), either directly or through a chain of operator functions such as $Filter$.\footnote{The $Annotated$ type requires at least 2 annotations and uses the first for type checkers.} As syntactic sugar, the schema from the first annotation is applied as a projection on the $Model$'s output.

In SQL, a query is preceded by a structured comment that specifies a schema object, by name, to be associated with the SQL result set. While this lacks support in the client environment in Step 1 (Fig.~\ref{fig:cloud}), it allows the control plane to associate contracts with SQL transformations at Step 2. Local static checking is therefore currently available only for Python nodes; SQL contracts first participate in validation during control-plane planning, while Step 3 enforcement at the Arrow boundary is language-independent across supported runtimes.

\subsection{Enforcement Semantics}

Contracts are enforced by ``fail fast'' behavior where errors are lifted to the earliest possible point in the execution lifecycle. When the user authors code, local type checkers can catch obvious mismatches immediately (Step 1). Next, before scheduling any execution, the control plane parses the metadata (Step 2) and validates that adjacent nodes compose correctly and that the boundary between tables in the catalog and the assets defined in the DAG is specified correctly. Finally, at the worker (Step 3), runtime checks validate that the physical data conforms to its specified schema before execution (input) and before any results are persisted (output), ensuring that late-discovered schema problems do not leak inconsistent state into storage.

\textbf{Enforcement at Steps 1 and 3} is relatively straightforward. Step 1 is entirely deferred to a local type checker as annotation syntax is purposefully designed to enable the use of standard type checkers. For example, $Table$ is defined in a way so that a type checker recognizes it as having the same interface as a $pyarrow.Table$, which prevents calling invalid methods on $adult\_lodgings$ in \texttt{g}.

At Step 3, a worker uses information from the control plane to validate that an output table (resulting from the transformation) has a correct schema as specified in the Bauplan project. For example, when \texttt{h} returns an Arrow table:

$$pl.DataFrame(f\_data).select(...).to\_arrow()$$

then an appropriate error message will be raised at runtime if schema validation fails:

  \begin{lstlisting}[
    language=Python,
    showstringspaces=false,
    columns=fullflexible,
    caption={Example schema errors},
    label=lst:runtime_err,
    basicstyle=\ttfamily\scriptsize,
    keepspaces=true,
    numbers=none
  ]
  # Some possible errors on column "Fare" after the transformation `h`
  Error: job failed: function failed due to user error:
  Schema contract validation failed on model [h]:
    - missing column "Fare" (expected double) (RuntimeTaskUserError)
    - column "Fare" has type int64, expected double (RuntimeTaskUserError)
    - unknown column "Fare" (has type int64) (RuntimeTaskUserError)
  \end{lstlisting}

Other table validations can then be performed, such as checking for null values (Section~\ref{sec:quality}). Finally, auxiliary logic may be executed, such as inserting column type casts when necessary.

\textbf{Step 2 Enforcement} involves a more complex analysis of the DAG types and how information and metadata flow across the DAG. We leverage a graph-based representation to reason about data, task, and infrastructure dependencies \cite{kuzu}, turning consistency checks into graph queries (e.g. do projections in the child node correctly map to its parent's output?). Only in Step 2 can the system link catalog information on source tables with the pipeline code, allowing deterministic type checks at the boundaries between DAG and catalog, and full reasoning over the data flowing from the lakehouse into the tasks being planned.

  \begin{lstlisting}[
    language=Python,
    showstringspaces=false,
    columns=fullflexible,
    caption={New node with type-cast},
    label=lst:type_cast,
    basicstyle=\ttfamily\scriptsize,
  ] 
  class TotalFares(TableSchema):
    """Simple schema for ticket fare data."""
    fare_total: Annotated[Int64, Required,
      Doc('Total ticket prices, rounded up to the nearest British pound.')
    ]
  
  @bauplan.python('3.13', pip={'polars': '1.37'})
  @bauplan.model(materialization_strategy='REPLACE')
  def fare_sum(
      tcosts: Annotated[Table[WomenByClass], Model('f')],
  ) -> Table[TotalFares]:
    """Compute the total fare for first- and second-class women passengers."""
    # Function body remains unchanged
    return (pl.DataFrame(tcosts)
              .select(fare_total=pl.col('Fare').sum().ceil())
              .to_arrow())
  \end{lstlisting}

With Listing~\ref{lst:type_cast} as a concrete example, we consider $Fare$ and how it flows from \texttt{titanic} to \texttt{f} to \texttt{fare\_sum}. In the catalog, \texttt{titanic} defines $Fare$ as a $double$, corresponding to $Float64$. In Listing~\ref{lst:schemas}, the schema $WomenByClass$ preserves $Fare$ as $Float64$ and adds documentation. The new node \texttt{fare\_sum} computes a sum, but wants the result to be rounded and returned as an $Int64$, as specified by its output type $TotalFares$. Before execution, the control plane determines that the output column $fare\_total$ must have type $Int64$ and includes this contract in the execution plan. At Step 3, the worker validates the produced Arrow column and, where permitted, applies the required cast before persistence.

\section{Capabilities}

SDK 2.0 is now in beta and will soon be included in our documentation: since our SDK is open-source, the client-side implementation is publicly available. In this section, we survey the consequences of this new design across four important dimensions: preventing the ``avoidable failures'', powering semantic reasoning through in-code documentation, enforcing data quality concisely and declaratively, and enabling engine composability without giving up contractual guarantees.

\subsection{Failure modes revisited}

With our enforcement semantics defined, we are now in a position to revisit the failure modes of SDK 1.0 and show how the new annotations address those challenges.

\begin{description}
    \item [columns dropped] If \texttt{g}'s declared output contract drops $tclass$ while \texttt{h} still requires it, the control plane rejects the DAG at Step 2. If \texttt{g}'s implementation omits $tclass$ while its declared contract still includes it, the worker rejects \texttt{g}'s output at Step 3, before persistence.
    \item [null values present] If \texttt{g} filters out $NULL$ values from its output, we can explicitly declare the non-null constraint and relate any violation to the node that defines it.
    \item [type change] If \texttt{g}'s declared type for $tclass$ changes from $Int64$ to $String$ while \texttt{h} still expects $Int64$, the control plane rejects the DAG at Step 2. If \texttt{g}'s implementation produces strings while its contract still declares $Int64$, the worker rejects \texttt{g}'s output at Step 3, before persistence.
    \item [shifting semantics] For a transformation using $tclass$ to disambiguate between \texttt{f} and \texttt{g}, we declare the desired lineage directly as \textit{WomenByClass['tclass']}
    where $WomenByClass$ is the schema for the output of \texttt{f}.
\end{description}

Each of the common failures now has a direct and concise solution.

\subsection{Semantic reasoning}
\label{sec:semanticlayer}

In line with the ``everything-as-code'' principle of modern data frameworks \cite{Tagliabue2023ReasonableSM}, SDK 2.0 couples transformation code with annotations describing the intended business meaning of the assets, as well as column-level metadata. In the \textit{write} path, everything-as-code guarantees that data semantics and data transformations are always ``in context'', which follows the best practices for AI-assisted coding and simplifies both human verification and ongoing code maintenance by coding agents.

\begin{figure}
\centerline{\includegraphics[width=0.45\textwidth]{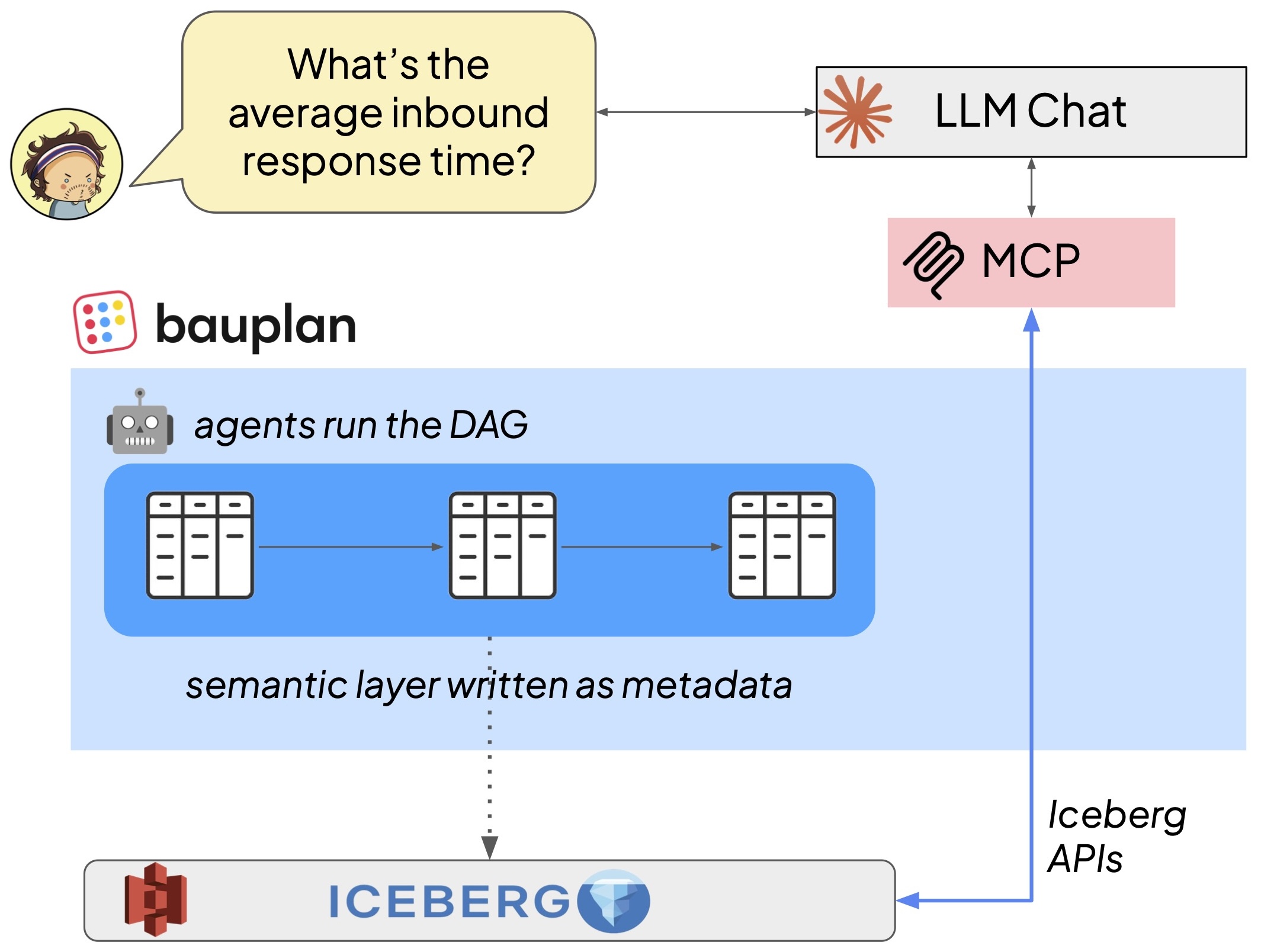}} 
\caption{A coding agent writes transformations and table metadata, which are persisted in Iceberg during a run. At \textit{read} time, an agent uses an MCP server exposing standard Iceberg metadata APIs to retrieve that information and put it into the LLM context for SQL generation.}
\label{fig:mcp}
\end{figure}

What about the \textit{read} path? The first step is therefore to make data semantics available outside the code base, without disconnecting them from the code that produced them. This is achieved at materialization time: when the materialization of assets occurs during a run, table and column annotations are persisted in versioned Iceberg table metadata associated with the materialized table state. When business users ask LLMs to translate English questions into queries (Figure~\ref{fig:mcp}), an agent using an MCP server that exposes Iceberg catalog and metadata APIs is able to retrieve asset descriptions and column information and put them into the LLM context for accurate SQL generation. In other words, annotations become part of the lakehouse exchange layer: any downstream system can recover the same semantic context, independently of what produced it. This turns metadata from local documentation into a composability mechanism across tools, engines, and agents.

Once again, the design stresses the \textit{outer} composability of Bauplan (how Bauplan and other data systems interact at predefined, clean interfaces) on top of the usual \textit{inner} composability (how Bauplan itself is built out of several modules exposed through vertical APIs).

\subsection{Data quality}
\label{sec:quality}

Types also give a principled handle on data quality checks without additional code -- such as manually writing \textit{expectations} in
SDK 1.0\footnote{\url{https://docs.bauplanlabs.com/concepts/expectations}} -- or additional tools. For example, Listing~\ref{lst:enum} illustrates an enumeration allowing declarative runtime checks without any user code to explicitly compare each value to $1$ or $2$. In particular, scalar constraints such as non-nullability and enumerated values can be checked directly on the Arrow buffer ``in-flight'', without the need for materialization and wasteful computation as is typical in \textit{dbt}-based setups.

\begin{lstlisting}[
    language=Python,
    showstringspaces=false,
    columns=fullflexible,
    caption={Supporting enum column types},
    basicstyle=\ttfamily\scriptsize,
    keepspaces=true,
    label=lst:enum
]
class WomenByClass(TableSchema):
    """Schema for women passengers with 1st or 2nd class tickets."""
    pid:    Annotated[Int64, Required, Doc('Unique passenger ID')]
    tclass: Annotated[Int64, Required,
        Enum('tclass', names=[('1st', 1), ('2nd', 2)]),
        Doc('Ticket class, filtered to 1st or 2nd class only.'),
    ]
\end{lstlisting}

Following the ``shift left'' philosophy, we envision a near future in which the SDK provides even further guarantees in the form of ``Dafny-style'' \textit{preconditions} and \textit{postconditions} for SQL nodes in a DAG. As an example, consider an aggregation-type node such as \texttt{SELECT col1, COUNT(*) FROM table GROUP BY ALL}: static analysis alone guarantees that \textit{if} no nulls are in \texttt{table.col1}, no new nulls will be created by this transformation -- in turn, if we knew through annotation and type inference that \texttt{table.col1} \textit{is} indeed not null, we could conclude via \textit{modus ponens} that no nulls will be returned at the end.

\subsection{Composability}
\label{sec:compo}

Through its \textit{connector} semantics\footnote{\url{https://docs.bauplanlabs.com/integrations/warehouses-lakehouses/snowflake-outbound}}, Bauplan supports replacing its proprietary sandboxes with alternative runtimes that implement the platform's execution interface. Because contracts are expressed over Arrow-compatible types, any such runtime that consumes and produces Arrow data can enforce the same input and output boundaries, even when the transformation engine changes. Further, annotations may be persisted in an Iceberg catalog and propagated between independent data pipelines or shared with other data platforms and subsystems (Section~\ref{sec:semanticlayer}).

As the composable data systems community expands, we envision independent platforms interpreting a common subset of these annotations, allowing contracts and semantic metadata to propagate across engines through Arrow and Iceberg interfaces.

\section{Related work}

Popular DAG frameworks provide partial forms of asset contracts. \textit{Dagster}~\cite{web:dagster} provides Python-centric asset checks, while \textit{dbt} model contracts~\cite{web:dbt} are limited to SQL models and perform a build-time preflight check of each model's query output against its YAML declaration. Unlike Bauplan, they do not type-check contracts across SQL and Python nodes or compose pipeline contracts against the current lakehouse catalog state before execution. 

Python libraries such as \textit{Pandera}~\cite{web:pandera} and \textit{Patito}~\cite{web:patito} similarly provide class-based schemas and runtime validation for DataFrames. Bauplan differs by attaching contracts to multi-language DAG boundaries, checking their composition against catalog state before execution, and enforcing them through a shared Arrow boundary.

The extensive use of $Annotated$ and the Pythonic DAG syntax are heavily inspired by Pydantic’s widely understood class-based annotation pattern \cite{gh:pydantic}. Separately, the idea of column lineage and its benefits is based on the analysis used in \textit{SqueezeCache} for ``squeezing'' \cite{web:squeezecache}. Our usage of lineage is currently more generic (closer to dataflow).

Pythonic data-quality tools such as \textit{Great Expectations}~\cite{web:greatexpectations} provide a rich framework for validating data assets through expectations and validation suites. However, they add a separate validation layer: users must adopt new dependencies, learn a distinct API, and write quality checks in addition to the transformation code. Our approach instead makes common data-quality constraints part of the table type itself. Nullability, enums, and column types are declared once in the schema and then interpreted uniformly across the client, control plane, and data plane.

\section{Conclusion}

We presented Bauplan SDK 2.0, a contract-oriented extension of our lakehouse programming model. Starting from common failures in multi-language pipelines, we introduced DAG types: lightweight schema objects that attach column types, constraints, documentation, and lineage to pipeline inputs and outputs. Bauplan interprets these annotations across the execution lifecycle, enabling earlier validation in the client, stronger contract composition in the control plane, and runtime checks before invalid results are persisted. The result is a portable contract layer for composable data systems, where correctness and semantics live next to the code that defines the data flow.

Our next step is to push these checks further left. We plan to develop a Bauplan Language Server Protocol (LSP) implementation that brings schema validation and actionable diagnostics directly into the IDE. This would move more reasoning client-side, reduce feedback latency, and enable even faster and safer agentic exploration over production data.

\bibliographystyle{ACM-Reference-Format}
\bibliography{sample}

\end{document}